\documentstyle[epsf]{mn}

\title[Possible detection of two giant extrasolar planets orbiting the eclipsing polar UZ Fornacis]{Possible detection of two giant extrasolar planets orbiting the eclipsing polar UZ Fornacis\thanks{Based on observations made with the Southern African Large Telescope (SALT)}}

\author[Stephen B. Potter et al.]  {Stephen B. Potter$^{1}$\thanks{sbp@saao.ac.za}, Encarni
  Romero--Colmenero$^{1}$, Gavin Ramsay$^{2,6}$, \and Steven Crawford$^{1}$,  Amanda Gulbis$^{1}$, Sudhanshu Barway$^{1}$, Ewald Zietsman$^{1,4}$, 
  \and Marissa Kotze$^{1,5}$, David A. H.  Buckley$^{1}$,  Darragh O'Donoghue$^{1}$,  \and O. H. W. Siegmund$^{3}$,  J. McPhate$^{3}$, 
   B. Y. Welsh$^{3}$ and John Vallerga$^{3}$ \\
  $^{1}$South African Astronomical Observatory, PO Box 9,
  Observatory 7935, Cape Town, South Africa \\
  $^{2}$Armagh Observatory, College Hill, Armagh BT61 9DG \\
  $^{3}$Experimental Astrophysics Group, Space Sciences Laboratory, University of California Berkeley, CA 94720, USA \\
   $^{4}$Department of Mathematical Sciences, The University of South Africa, PO Box 392, UNISA, 0003, South Africa \\
  $^{5}$Astronomy Department, Astrophysics, Cosmology and Gravity Centre (ACGC), University of Cape Town, Rondebosch 7701, South Africa \\
  $^{6}$Mullard Space Science Laboratory, University College London, Holmbury St Mary, Dorking, Surrey RH5 6NT, UK
}

\date{}

\begin{document}

\maketitle

\begin{abstract}

%We have detected departures in the eclipse times of UZ For from a
%simple quadratic ephemeris of up to ∼ 60s. The de- partures are
%suggestive of two periodicities of ∼ 15.7 and ∼ 5.24 years. The two
%favoured mechanisms, to drive the periodicities, are either two giant
%extrasolar planets as com- panions to the binary or a magnetic cycle
%mechanism (e.g. Applegate’s mechanism) of the secondary star. However,
%Ap- plegate’s mechanism would require the entire radiant energy output
%of the secondary to drive such a mechanism and therefore would seem to
%be the least likely of the two, bar- ing any further refinements in
%the effect of magnetic fields (e.g. Lanza 1998). The strong indication
%of a 3:1 period ratio would seem to add more weight to the two planet
%model.

We present new high-speed, multi-observatory, multi-instrument
photometry of the eclipsing polar UZ For in order to measure precise
mid-eclipse times with the aim of detecting any orbital period
variations. When combined with published eclipse times and archival
data spanning $\sim $27 years, we detect departures from a linear and
quadratic trend of $\sim$60 s. The departures are strongly suggestive
of two cyclic variations of 16(3) and 5.25(25) years.  The two
favoured mechanisms to drive the periodicities are either two giant
extrasolar planets as companions to the binary (with minimum masses of
6.3(1.5)$M_{Jup}$ and 7.7(1.2)$M_{Jup}$) or a magnetic cycle mechanism
(e.g. Applegate's mechanism) of the secondary star.  Applegate's
mechanism would require the entire radiant energy output of the
secondary and would therefore seem to be the least likely of the two,
barring any further refinements in the effect of magnetic fields
(e.g. those of Lanza et al.).  The two planet model can provide
realistic solutions but it does not quite capture all of the eclipse
times measurements. A highly eccentric orbit for the outer planet
would fit the data nicely, but we find that such a solution would be
unstable.  It is also possible that the periodicities are driven by
some combination of both mechanisms. Further observations of this
system are encouraged.

\end{abstract}


\section{Introduction}

%The standard picture of a cataclysmic variable (CV) is a binary system
%consisting of a Roche lobe filling red dwarf (known as the secondary
%or the donor star) and an accreting white dwarf (the primary). CVs
%have orbital periods of typically a few hours, and mass transfer is
%caused by angular momentum loss - see e.g. Warner (1995) for a review
%of cataclysmic variables.  

Approximately 20\% of the known cataclysmic variables (CVs, see the
catalogue of Ritter \& Kolb 2003) are polars, where the
primary white dwarf has a sufficiently strong magnetic field to lock
the system into synchronous rotation with the red dwarf secondary and
to prevent completely the formation of an accretion disc.

The material from the secondary overflowing the Roche lobe
initially falls towards the white dwarf following a ballistic
trajectory until, at some distance from the white dwarf, the magnetic
pressure overwhelms the ram pressure of the ballistic stream. From
this point on the accretion flow is confined to follow the magnetic
field lines of the white dwarf. The now supersonic accreting material
suddenly becomes sub-sonic at a shock region, which forms at some
height above the white dwarf surface. The shock-heated material
reaches temperatures of $\sim 10-50$ keV and is therefore ionised. The
hot plasma cools by X-ray cooling, in the form of bremsstrahlung
radiation. With sufficiently strong magnetic fields we find also
cyclotron cooling, in the form of optical/infrared cyclotron radiation
(e.g. ST LMi: Imamura, Steiman-Cameron \& Wolff, 2000, and Campbell,
Harrison, Mason, Howell \& Schwope, 2008). See e.g. Warner (1995) for
a review of CVs and Cropper (1990) and Patterson (1994) for reviews of
magnetic CVs.

%hese are further sub-divided into two subtypes, namely
%intermediate polars (IPs) and polars, depending on the strength of the
%magnetic field of the white dwarf and the degree of synchronism
%between the white dwarf spin and the binary orbit- see the reviews
%given by Cropper (1990) and Patterson (1994).

UZ For is one of 15 known eclipsing polars and was discovered with
{\it EXOSAT} (EXO 033319-2554.2) as a serendipitous X-ray source
(Giommi et al. 1987; Osborne et al. 1988). Extensive followup
observations at multi-wavelengths established a $\sim $126.5-min
orbital period, one or two accretion spots depending on accretion
state, with magnetic fields of $\sim$53 MG and $\sim$48 MG and a white
dwarf mass of $\sim$$0.7$$\rm M_{\odot}$ (Beuermann, Thomas \& Schwope
1988; Berriman \& Smith 1988; Ferrario et al. 1989; Bailey \& Cropper
1991). The eclipses of the accretion spots in UZ For are particularly
rapid at 1-3 s and can only be resolved with high speed
photometry. Bailey \& Cropper (1991) were able to resolve the eclipse
of the white dwarf photosphere during a low state and Perryman et
al. (2001) were able to resolve two accretion spots during a higher
accretion state. These distinct rapid photometric transitions are
ideal for making accurate timing measurements and therefore searching
for any long term period variations in UZ For. Some of the
above-mentioned authors have combined their observations with previous
eclipse measurements in order to obtain accurate eclipse
ephemerides. In general, significant residuals were seen in the O-Cs
(Observed - Calculated) of the orbital period, but no overall trend
had been detected (e.g. Perryman et al. 2001). More recently, Dai et
al. (2010) claim the existence of a third body orbiting UZ For in
order to explain the O-C. However, their singular new eclipse
measurement and subsequent derived orbital parameters are grossly
incompatible with all of our new observations spanning $\sim$10 years.

Nevertheless, recent results of long term studies of some CV related
objects are beginning to show trends. Parsons et al. (2010) presented
high-speed ULTRACAM photometry of 8
post-common-envelope-binaries. They detect significant departures from
linearity in some of these systems and suggest magnetic braking or a
third body as possible mechanisms to drive the O-Cs. High precision
eclipse measurements of the post-common envelope binary NN Ser
(Beuermann et al. 2010a) shows strong evidence for two additional
bodies superposed on the binary's linear ephemeris. Significant and
complicated departures from a linear ephemeris have also been seen in
the eclipsing polar HU Aqr (Schwarz et al. 2009). They find that
neither a sinusoidal nor a quadratic ephemeris are sufficient to
describe their O-C departures, thus more eclipse observations over the
next few years will be needed in order to refine the ephemerides. Qian
et al. (2010) discovered that the O-C curve of the eclipsing polar DP
Leo shows a cyclic variation with a period of 23.8 years. They claim
that this is as a result of a giant extrasolar planet orbiting DP Leo,
recently refined by Beuermann et al (2010b).

%Past quadratics etc in UZ For ... non-consistent time units etc and
%not sufficiantly long time base.

Here we present new high-speed {\it HIPPO, BVIT, SALTICAM} and {\it
  UCTPOL} photometry of UZ For, spanning 10 years, and use these
observations to determine accurate mid-eclipse times of the main
accretion spot in UZ For. We combine these with previous mid-eclipse
times, that we either measure from archival data or extract from the
literature, and analyse for any period variations in UZ For. Note that
adding these newer data to timing from the literature gives a baseline
of 27 years.

\section{Observations}

All of the eclipse times extracted from the literature were published
as Heliocentric Julian Dates (HJD). We have assumed that the
Coordinated Universal Time (UTC) system was used in all cases as this
was not explicitly stated in any of the publications. We re-corrected
all times for the light travel time to the barycenter of the solar
system, converted to the barycentric dynamical time system (TDB) and the
times are listed (table \ref{tab:observations}) as Barycentric Julian
Date (BJD; see Eastman, Siverd \& Gaudi 2010 for achieving accurate
absolute times and time standards). By doing this we have removed any
timing systematics, particularly due to the unpredictable accumulation
of leap seconds with UTC, and effects due to the influence of
primarily Jupiter and Saturn when heliocentric corrections only are
applied. We either calculate or re-calculate appropriate errors
depending on the S/N and time resolution at the time of the spot
ingress and egress. Table \ref{tab:observations} also lists the
eclipse width of the accretion spot and the observatory/instrument
used.

All of our new observations were also converted to BJD. We note that
our new ground based instruments were synchronised to GPS to better
than a milli-second. Given the high-speed nature of these instruments,
their timing accuracies have been verified through simultaneous
multi-instrument observations. The remaining space observatories have
documented reports on the performance of their on-board clocks.

\subsection{\it Eclipse times from the literature}

The earliest UZ For eclipse measurements were made using {\it EXOSAT}
and published by Osborne et al. (1998). The data are of poor time
resolution but are at a sufficiently early epoch to provide
constraints for model fitting. Beuermann et al. (1988) and Ferrario et
al. (1989) observed multiple eclipses spectrophotometrically. These
are also of very low time resolution, however the combination of
multiple eclipses provides usable data. Allen et al. (1989) presented
the first high speed photometry that could resolve the accretion
spot. A typographical error in the eclipse time was corrected by
Ramsay (1994). This was soon followed with more high quality optical
low-state photometry by Bailey \& Cropper (1991) and high-state
photometry by Imamura \& Steiman-Cameron (1998) and {\it EUVE} light
curves by Warren et al. (1995). An additional high quality STJ eclipse
was also published by de Bruijne (2002) and three more, with the same
instrument by Perryman et al. (2001). From the latter data we were
able to obtain the raw observations and re-measure and confirm the
eclipse times.

%{\it Bailey and Cropper} We are unable to retreive and reduce this
%data set ourselves. The assumption is that B+C eclipse times are in
%UTC. Also, whatever software they used to calculate HJD assumed input
%was UTC and output is UTC. I.e. that no other conversions (e.g. from
%UTC to TT) were done in their HJD calculation software etc. Using luis
%star software to reproduce HJD and hence get JD: 47829.18421D HJD(utc)
%Bailey and Cropper = 47829.18008 JD(utc) Now use $calc_bjdfromutc$ to
%get BJD(tdb) = 47829.18486375D Which agrees with list from Gav to 0.32
%secs

\subsection{\it ROSAT (1991)}

%/Users/sbp/uzfor/rosat
%plot_rawBJD_tdb

Observations were retrieved from the HEASARC archive and events were
extracted using an aperture centered on the source and also a
background region. The resulting light curves were subtracted after
appropriate scaling from the differing areas. The observations spanned
$\sim$1 day and multiple eclipses were covered. The reduced data were
binned and folded on the orbital period in order to produce a single
eclipse light curve from which the eclipse was measured.

%Observations were retrieved from the {\it HEASARC} archive and reduced
%using the pipeline software {\bf GAV?}.  

%Can see eclipse of spot 1  only, cannot see not spot 2.

%ingress=2448482+      0.72526423 +-   6.1505803e-05
%egress=2448482+      0.73078423 +-   6.1505803e-05
%mid=2448482+      0.72802423
%width=    0.0055200000

%\subsection{\it STJ Photometry 1999}

%Perryman et al. 2001

%Gavin re-reduced the observations and provided light curves in JD(utc).
%We used $calc_bjdfromutc.pro$ to calculate BJD(tdb).

%1999 data covered 3 eclipses.

%All data folded NOT binned in order to calc mid eclipse times.
%Can see eclipse of spot 1 and spot 2.

%Spot 1: \\
%ingress=2451522+      0.43002957 +-   2.6359630e-05 \\
%egress=2451522+      0.43542957 +-   2.6359630e-05 \\
%mid=2451522+      0.43272957 \\
%width=    0.0054200000 \\
%Spot 2:\\
%$sm_ingress=2451522+      0.43035072 +-   4.3932716e-05 \\
%sm_egress=2451522+       0.43497957 +-   6.1505803e-05 \\
%$
%2000 observations recovered from literature

%From the oringinal et al perryman et al scam spie paper

%Reading off their plot:

%04:30:41–05:00:41

%   ingress 04 30 41 +642s    egress 04 30 41 +1109s     
%=  ingress 04 41 23          egress 04 49 10  = total eclipse = 467s

%mid eclipse = 04 41 23 + 234s = 04 44 77 = 04 45 17 UTC

%2000 10 4 04 45 17 UTC = 2451821.70239393 BJD(TDB)

%Use same error estimates as above for 1999.

%Second pole not so visible ... need to be able to  expand plot.

\subsection{\it HST FOS (1992)}

UZ For was observed by HST with FOS on 11th June 1992 in two
consecutive, RAPID mode observations consisting of 925 spectra each
with 1.64s exposure times. This data set was originally published in
Stockman \& Schmidt (1996), but their mid-eclipse times were not
quoted. Therefore the HST data products from this observation were
downloaded from the HST archive at the MAST. The flux- and
wavelength-calibrated individual spectra were extracted and the flux
summed-up between 1255 and 1518\AA (far UV) to create a
lightcurve. The far UV part of the spectrum was chosen as this seems
to have the least contribution from the accretion stream.  The
end-times for each spectra were obtained from the observation header
keywords and the group-delay-time subtracted in order to obtain times
of start of each exposure. Two consecutive eclipses were observed and
folded and binned into a single eclipse from which measurements were
made.

%The time units were in MJD(=JD-2450000.5), and converted to BJD(TDB). 

%These times dont yet look right ... MJD is not what it should be

%ingress=2448784+      0.71873428 +-   2.6359630e-05
%egress=2448784+      0.72410428 +-   3.5146173e-05
%mid=2448784+      0.72141928
%width=    0.0053700000

\subsection{\it EUVE 1993 and 1995 }

UZ For was observed with EUVE on the 18th November 1993 and on the
15th Jan 1995 for 102ks and 76ks respectively. These data were
retrieved from the STScI archive and reduced following the recipe from
http://archive.stsci.edu/euve/. Lightcurves were produced using the
xray.xtiming package in IRAF with a 1s time resolution. For each
observation, multiple eclipses were covered which were folded and
binned into two single eclipses from which measurements were made.

%The ouptut times were converted from MJD
%(=JD-2450000.5) to JD(UT), and then to BJD(TDB).
%$calc_bjdfromutc.pro$

%ingress=2449310+      0.32986882 +-   4.3932716e-05
%egress=2449310+      0.33531882 +-   4.3932716e-05
%mid=2449310+      0.33259382
%width=    0.0054500000

%gav's conversion agrees to 1.7s 

%scanner pointings not in archive

%\subsection{\it EUVE 1995 }

%UZ For was observed with EUVE on the 15th Jan 1995. The 1995 data were
%retrieved from the STScI archive and reduced following the recipe from
%http://archive.stsci.edu/euve/

%Lightcurves were produced using the xray.xtiming package in IRAF with
%a 1s time resolution.  The ouptut times were converted from MJD
%(=JD-2450000.5) to JD(UT), and then to BJD(TDB) using
%$calc_bjdfromutc.pro$

%ingress=2449733+      0.40248733 +-   4.3932716e-05
%egress=2449733+      0.40768733 +-   4.3932716e-05
%mid=2449733+      0.40508733
%width=    0.0052000000

\subsection{\it UCTPOL 2002 and 2005}

One unfiltered and two BG39 filtered eclipses were obtained in 2002
with the University of Cape Town photo-polarimeter (UCTPOL) at 10
second time resolution. Two eclipse times were extracted: one from the
unfiltered and the second from the folded BG39 filtered
eclipses. Three unfiltered eclipses were obtained in 2005 at 10 and 1
second time resolution. Two eclipse times were extracted. On both
occasions, simultaneous linear and circular polarimetric observations
were also made and reduced as in Cropper (1985).

%1 eclipse .. 2 poles visible, clear filter \\

%ingress=2452493+      0.60634552 +-   7.0292346e-05 \\
%egress=2452493+      0.61177052 +-   7.0292346e-05 \\
%mid=2452493+      0.60905802 \\
%width=    0.0054250000 \\
%$sm_ingress=2452493+      0.60669419 +-   7.0292346e-05 \\
%sm_egress=2452493+      0.61133743 +-   7.0292346e-05 \\
%$

%2 eclipses .. folded ... 2 poles visible, bg39 filter \\

%ingress=2452494+      0.57291318 +-   3.5146173e-05 \\
%egress=2452494+      0.57833818 +-   3.5146173e-05 \\
%mid=2452494+      0.57562568 \\
%width=    0.0054250000 \\
%$sm_ingress=2452494+      0.57326184 +-   7.0292346e-05 \\
%sm_egress=2452494+      0.57792441 +-   7.0292346e-05 \\
%$

%\subsection{\it UCTPOL 2005}

%3 eclipses .... 2 poles viisble \\

%ingress=2453405+      0.29795053 +-   3.5146173e-05 \\
%egress=2453405+      0.30337553 +-   8.7865433e-06 \\
%mid=2453405+      0.30066303 \\
%width=    0.0054250000 \\
%$sm_ingress=2453405+      0.29829919 +-   3.5146173e-05 \\
%sm_egress=2453405+      0.30294724 +-   2.6359630e-05 \\
%$

%ingress=2453407+      0.31886188 +-   8.7865433e-06 \\
%egress=2453407+      0.32428688 +-   8.7865433e-06 \\
%mid=2453407+      0.32157438 \\
%width=    0.0054250000 \\
%$sm_ingress=2453407+      0.31921055 +-   2.6359630e-05 \\
%sm_egress=2453407+      0.32385379 +-   2.6359630e-05 \\
%$

%ingress=2453408+      0.28537331 +-   8.7865433e-06 \\
%egress=2453408+      0.29079831 +-   8.7865433e-06 \\
%mid=2453408+      0.28808581 \\
%width=    0.0054250000 \\
%$sm_ingress=2453408+      0.28572198 +-   2.6359630e-05 \\
%sm_egress=2453408+      0.29036522 +-   2.6359630e-05 \\
%$

\subsection{\it XMMOM 2002}

%{\it XMMOM} observations made in 2002, 2 orbits after the UCTPOL 2002
%observations. The observations extracted from the {\it HEASARC}
%archive and reduced using the standard pipeline software (GAV?). Three
%consequative eclipses were observed, unfortunately two missed the
%ingress. Nevertheless, upon folding and binning a good quailty eclipse
%profile was obtained and eclipse measurements taken.

Observations were made in fast-mode using the {\it XMM-Newton} Optical
Monitor (Mason et al 2001) between Aug 7 and Aug 8 2002; two orbits
after UCTPOL observations. The UVW1 filter was used (center wavelength
2910\,\AA, FWHM 500\AA) and the data were reduced using {\tt omfchain}
running under {\tt SAS} v9.0. Although three consecutive eclipses
were observed, two had incomplete coverage. Nevertheless, upon folding
and binning, a high signal-to-noise eclipse profile was obtained and
eclipse measurements taken.

%All data folded and binned in order to calc mid eclipse times.

%ingress=2452494+      0.83642110 +-   5.2719260e-05
%egress=2452494+      0.84197110 +-   8.7865433e-05
%mid=2452494+      0.83919610
%width=    0.0055500000

\subsection{\it SWIFT 2005}

%Data reduced using blah blah blah (Gav?)

%UZ For was visited 3 times over 4 days in which 3 eclipses were
%observed within 1 day of the UCTPOL 2005 observations.  The data were folded
%and binned into a single eclipse from which measurements were made.

Observations were made in event mode using the {\it SWIFT} UV Optical
Telescope (Roming et al 2005) between Feb 2 and Feb 6 2005. There were
a number pointings, some lasting a few 100 sec and others a few 1000
sec. The $U$ filter (center wavelength 3450\AA, FWHM 875\AA) and the
$V$ filter were used. Light curves were extracted using apertures
centered on the source (radius 3$^{''}$) and also a source free
background region with much larger aperture radius. The light curve
was generated by suitably scaling the size of the apertures. Two
eclipses were observed in full, one of which was simultaneous with the
UCTPOL observations.

%1 eclipse
%1/2 eclipse
%1 eclipse

%ingress=2453404+      0.33134192 +-   4.3932716e-05
%egress=2453404+      0.33674192 +-   6.1505803e-05
%mid=2453404+      0.33404192
%width=    0.0054000000

\subsection{\it SALTICAM 2007}
UZ For was observed with SALTICAM (O'Donoghue et al. 2006) on SALT on
12 November 2007. SALTICAM was in slot-mode configuration, allowing a
time resolution of 1s with no deadtime. The data were reduced using
the SALT slottools data reduction package (Crawford et al. 2010). One
eclipse of high time resolution and signal to noise was observed from
which measurements were made.

%... 1 pole visible only
%wd photosphere clearly visible  .... low state?

%The times of the start of each
%exposure where UT. These times were converted to BJD(TDB)
%$calc_bjdfromutc.pro$

%ingress=2454417+      0.33201170 +-   8.7865433e-06 \\
%egress=2454417+      0.33743170 +-   8.7865433e-06 \\
%mid=2454417+      0.33472170 \\
%width=    0.0054200000 \\

\subsection{\it BVIT 2009}
BVIT (Berkeley Visible Imaging Tube: Siegmund et al. 2008) is a
visible photon counting detector designed as a guest facility on the
SALT to provide very high time resolution ($<$25 nanoseconds) and high
signal to noise, full imaging photometry.  UZ For was observed during
a BVIT commissioning run on 25th January 2009 simultaneous with the
HIPPO on the 1.9m telescope of the South African Astronomical
Observatory.  The data were extracted making use of the IDL data
reduction software developed by the instrument team and binned into
0.5s bins. One eclipse of high time resolution and s/n was observed
from which measurements were made.

% ... 2 poles visible \\ wd
%photosphere clearly visible \\ Accretion stream visible late into
%eclipse, similar to HST UV photometry (Stockman \& Schmidt 1996).

%ingress=2454857+      0.36209267 +-   8.7865433e-06 \\
%egress=2454857+      0.36751767 +-   8.7865433e-06 \\
%mid=2454857+      0.36480517 \\
%width=    0.0054250000 \\
%$sm_ingress=2454857+      0.36241319 +-   1.3179815e-05 \\
%sm_egress=2454857+      0.36708938 +-   1.5815778e-05 \\
%$

\subsection{\it HIPPO January 2009, September 2010,  October 2010 and November 2010 }

Unfiltered photo-polarimetric observations were made with the HIPPO
(HI speed Photo-POlarimeter: Potter et al. 2010) on four separate
occasions and reduced as in Potter et al. (2010). The single eclipse
in January 2009 was observed simultaneously with the BVIT
observations. Multiple eclipses were observed on the other occasions
which were folded and binned from which measurements were made.

%1eclipse simult BVIT 2 poles visible \\

%2 consequative nights, 1 eclipse each, folded and binned

%3 eclipses from 2 nights combined ... perhaps see 2nd pole?

%ingress=2454857+      0.36209600 +-   8.7865433e-06 \\
%egress=2454857+      0.36752100 +-   1.7573087e-05 \\
%mid=2454857+      0.36480850 \\
%width=    0.0054250000 \\
%$sm_ingress=2454857+      0.36244466 +-   3.5146173e-05 \\
%sm_egress=2454857+      0.36708791 +-   1.7573087e-05 \\
%$

%\subsection{\it HIPPO Sept, Oct 2010 } %\subsection{\it HIPPO Sept 2010 }

%ingress=2455478+      0.48312116 +-   1.7573087e-05
%egress=2455478+      0.48854116 +-   2.6359630e-05
%mid=2455478+      0.48583116
%width=    0.0054200000

%ingress=2455450+      0.54191582 +-   2.6359630e-05
%egress=2455450+      0.54732582 +-   2.6359630e-05
%mid=2455450+      0.54462082
%width=    0.0054100000

%\subsection{\it EUVE 1998 aug}
%looks flat ... cant see eclipse??? 2 targets on image??

%\subsection{\it EUVE 1998 sept}
%does not look right ... cnts get high at the end of every run

%2 consequative eclipses folded and binned
%ingress=2455506+      0.42432435 +-   1.7573087e-05
%egress=2455506+      0.42974435 +-   2.6359630e-05
%mid=2455506+      0.42703435
%width=    0.0054200000

\begin{figure*}
\epsfxsize=16cm
%\begin{figure}
%\epsfxsize=8cm
%\epsffile{eclipses.ps} 
\epsffile{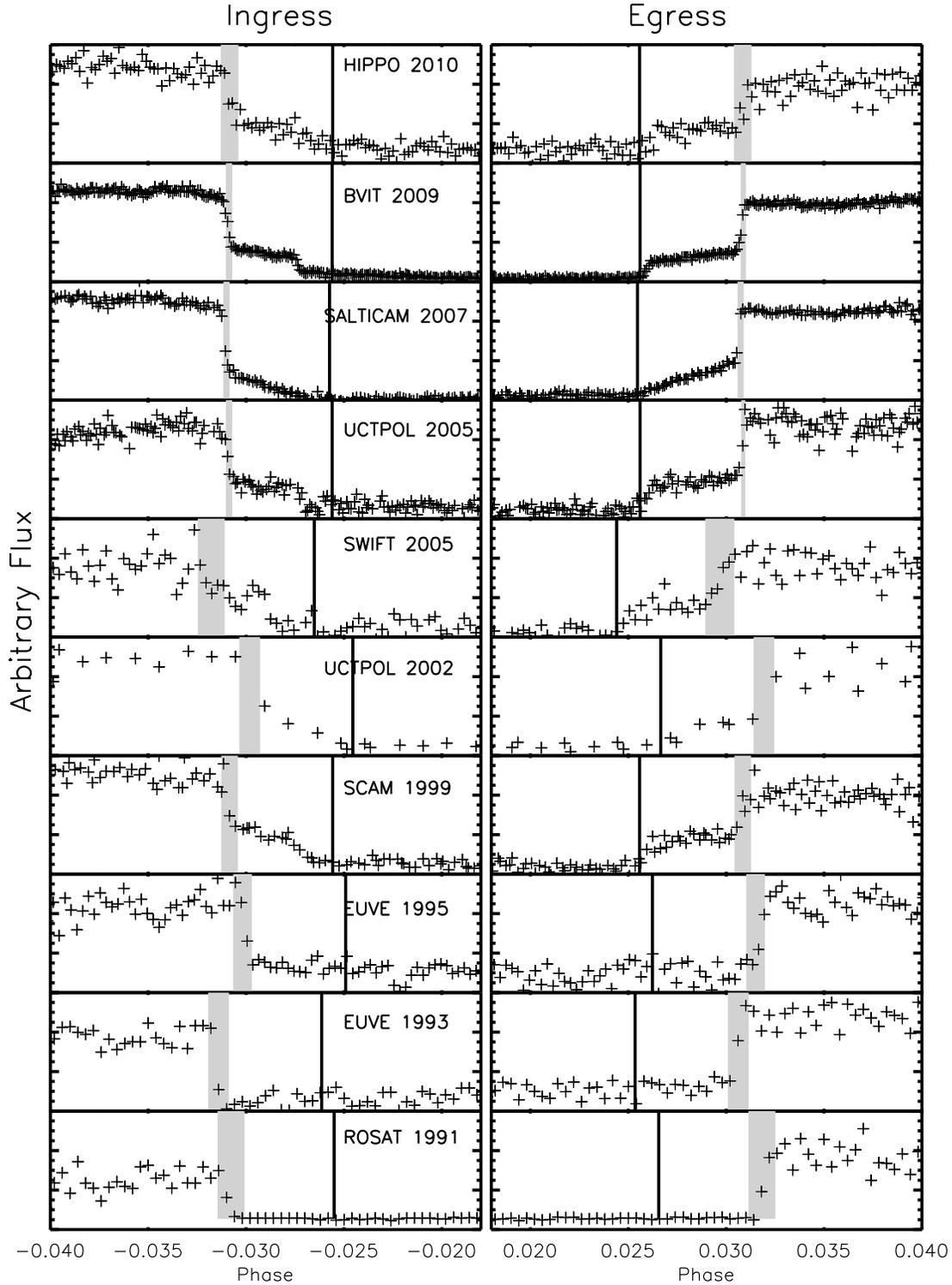} 
\caption{A sample of our new eclipse observations phased on our new
  ephemeris (section 4.2) Vertical grey bars indicate ingress and
  egress of the main accretion spot. Solid vertical bars indicate
  times of white dwarf ingress and egress assuming a duration of
  40s. }
\label{eclipses}
\end{figure*}

% Note that we have not included the eclipses of Dai et
%  al. (2010) as their O-Cs are over 300s compared to our linear ephemeris. We
%  believe that either their measurement and/or time standard conversion are
%  in error.

\begin{table*}
\begin{center}
\caption{Mid-eclipse times of the main accretion spot of UZ For. BJD$_{TDB}$ is the
  Barycentric Julian Date in the barycentric dynamical time
  system. Times have also been barycentrically corrected.  $^{1}$de
  Bruijne et al. (2002), $^{2}$Perryman et al. (2001), $^{3}$Imamura
  \& Steiman-Cameron (1998), $^{4}$Warren et al. (1995), $^{5}$Ramsay
  (1994), $^{6}$Bailey \& Cropper (1991), $^{7}$Allen et al. (1989),
  $^{8}$Ferrario (1989), $^{9}$Beuermann et al. (1988), $^{10}$Osborne
  et al (1988).
 {\label{tab:observations}}} \vspace{0.2cm}
\centerline{
\begin{tabular}{|r|l|l|c|c|} \hline
Cycle  & BJD$_{TDB}$+2400000 & $\Delta$BJD$_{TDB}$ & Width(sec) & Observatory/Instrument \\ \hline
23913 & 55506.42703435& 0.00001    & 468(2) & 1.9m/HIPPO \\
23595 & 55478.48583116& 0.00001    & 468(2) & 1.9m/HIPPO \\
23277 & 55450.54462082& 0.00001  & 467(2) & 1.9m/HIPPO \\
16526 & 54857.36480850& 0.00001    & 469(2) & 1.9m/HIPPO \\
16526 & 54857.36480517& 0.0000086  & 469(1) & SALT/BVIT \\
11518 & 54417.33472170& 0.0000086  & 468(1) & SALT/SALTICAM \\
34    & 53408.28808581& 0.0000086  & 469(1) & 1.9m/UCTPOL \\
23    & 53407.32157438& 0.00001  & 469(2) & 1.9m/UCTPOL \\
0     & 53405.30066303& 0.000035   & 469(3) & 1.9m/UCTPOL \\
-11.0 & 53404.33404192& 0.00006    & 467(4) & SWIFT \\
-10362& 52494.83919610& 0.000087   & 479(8) & XMM OM \\
-10365& 52494.57562568& 0.000035   & 469(3) & 1.9m/UCTPOL \\
-10376& 52493.60905802& 0.00007    & 469(6) & 1.9m/UCTPOL \\
-18023& 51821.70239393& 0.00001    & 467(2) & WHT/SCAM 2000$^{1}$ \\
-21360& 51528.49543399& 0.00002    & 468(2) & WHT/SCAM 1999$^{2}$ \\
-21361& 51528.40757990& 0.00002    & 468(2) & WHT/SCAM 1999$^{2}$ \\
-21429& 51522.43272958& 0.00002    & 468(2) & WHT/SCAM 1999$^{2}$ \\
-38508& 50021.779388&   0.00005    & & CTIO 1m/Photometer$^{3}$ \\
-38543& 50018.704108&   0.00005    & & CTIO 1m/Photometer$^{3}$ \\
-41537& 49755.634978&   0.00005    & & CTIO 1m/Photometer$^{3}$ \\
-41538& 49755.547148&   0.00005    & & CTIO 1m/Photometer$^{3}$ \\
-41560& 49753.614028&   0.00005    & & CTIO 1m/Photometer$^{3}$ \\
-41571& 49752.647568&   0.00005    & & CTIO 1m/Photometer$^{3}$ \\
-41790& 49733.40501704& 0.00004    & 467(4) & EUVE\\
-46605& 49310.33259382& 0.00003    & 471(4) & EUVE\\
-46988& 49276.680055&   0.00004    & & EUVE$^{4}$\\
-52587& 48784.72141928&  0.00003   & 463(4) & HST \\
-56024& 48482.72808573&  0.0001     & 477(5) & ROSAT$^{5}$ \\
-63462& 47829.18486375& 0.00003    & & AAT$^{6}$ \\
-63474& 47828.130520&   0.00003    & & AAT$^{6}$ \\
-63476& 47827.954780&   0.00003    & & AAT$^{6}$ \\
-67915& 47437.919920&   0.00003    & 466.5(2.5)& 2.3m Steward Obs.$^{7,5}$ \\
-71248& 47145.064339&   0.0002     & & AAT$^{8}$ \\
-71451& 47127.227739&   0.0002     & & AAT$^{8}$ \\
-71452& 47127.139439&   0.0002     & & AAT$^{8}$ \\
-71786& 47097.792559&   0.0002     & & ESO/MPI 2.2m$^{9}$ \\
-71821& 47094.717359&   0.0002     & & ESO/MPI 2.2m$^{9}$ \\
-71857& 47091.554239&   0.0002     & & ESO/MPI 2.2m$^{9}$ \\
-71868& 47090.587789&   0.0002     & & ESO/MPI 2.2m$^{9}$ \\
-71889& 47088.742549&   0.0002     & & ESO/MPI 2.2m$^{9}$ \\
-79193& 46446.973809&   0.00016    & & EXOSAT$^{10}$ \\
-89206& 45567.177597&   0.00016    & & EXOSAT$^{10}$ \\
\hline
\end{tabular}
}     
\end{center}
\end{table*}

\section{Results}

\begin{table*}
\caption{ Mid-eclipse ephemerides of the main accretion spot of UZ For
  and corresponding planet model parameters. Ephemeris parameters
  correspond to the {\it representative} solution (Fig~\ref{per_per})
  and are rounded off to the 1 sigma errors.  The planet parameter
  errors were calculated using the range in parameter space of
  possible solutions and not the smaller 1 sigma errors of any one
  fit. Minimum planet masses are listed assuming
  coplanearity. $M_{3,4,fnc}$ is the mass function. The combined mass of the
  primary and secondary stars are assumed to be $0.84 \rm M_{\odot}$.
  {\label{tab:planets}}}% \vspace{0.2cm}
\begin{tabular}{|l|l|l|l} \hline 
{\bf Quadratic term:}       & $T_{0}$ = 2453405.30086(3) d                &                                           &   \\ 
                            & $P_{bin}$ = 0.087865425(2) d                & {\bf Planet }                             &    \\ 
                            & $A$ = $-7(2) 10^{-14}$                      & {\bf Parameters: }                        & \\ \hline
{\bf 1st Elliptical term:}  & $\upsilon_{3}=($E$+T_{3})f_{3} $             & $M_{3,fnc}=2.9(1.1) 10^{-7}\rm M_{\odot} $  & \\ 
                            &$T_{3}$  = 60383(416) (binary cycle)         & $M_{3,Jup}=6.3(1.5)$                      & \\ 
                            &$f_{3}$  = 0.000098(3) (cycles/binary cycle) & $P_{3} = 16(3)$ years                     & \\ 
                            & $\varpi_{3}=0.85(5)$                        & $a_{3} = 5.9(1.4)$ AU                     & \\
                            & $K_{bin,(3)}$ = 0.00025(2) d                 & $a_{1,2} = 0.042(1)$ AU                   & \\
                            & e = 0.04(5)                                 &                                          &\\ \hline
{\bf 2nd Elliptical term:}  & $\upsilon_{4}=($E$+T_{4})f_{4} $             & $M_{4,fnc}=5.3(5)\ 10^{-7}\rm M_{\odot} $   & \\
                            & $T_{4}$ = 4833(215)  (binary cycle)         & $M_{4,Jup}=7.7(1.2)$                      & \\
                            & $f_{4}$ = 0.000288(2) (cycles/binary cycle) & $P_{4} = 5.25(25)$ years                  & \\
                            & $\varpi_{4} = 1.20(6)$                      & $a_{4} = 2.8(5)$ AU                       & \\ 
                            & $K_{bin,(4)}$ = 0.000141(6) d                & $a_{1,2} = 0.025(1)$ AU                   & \\
                            & e = 0.05(5)                                 &                                          &\\ \hline
\end{tabular}
\end{table*}

\begin{figure*}
\epsfxsize=16cm
\epsffile{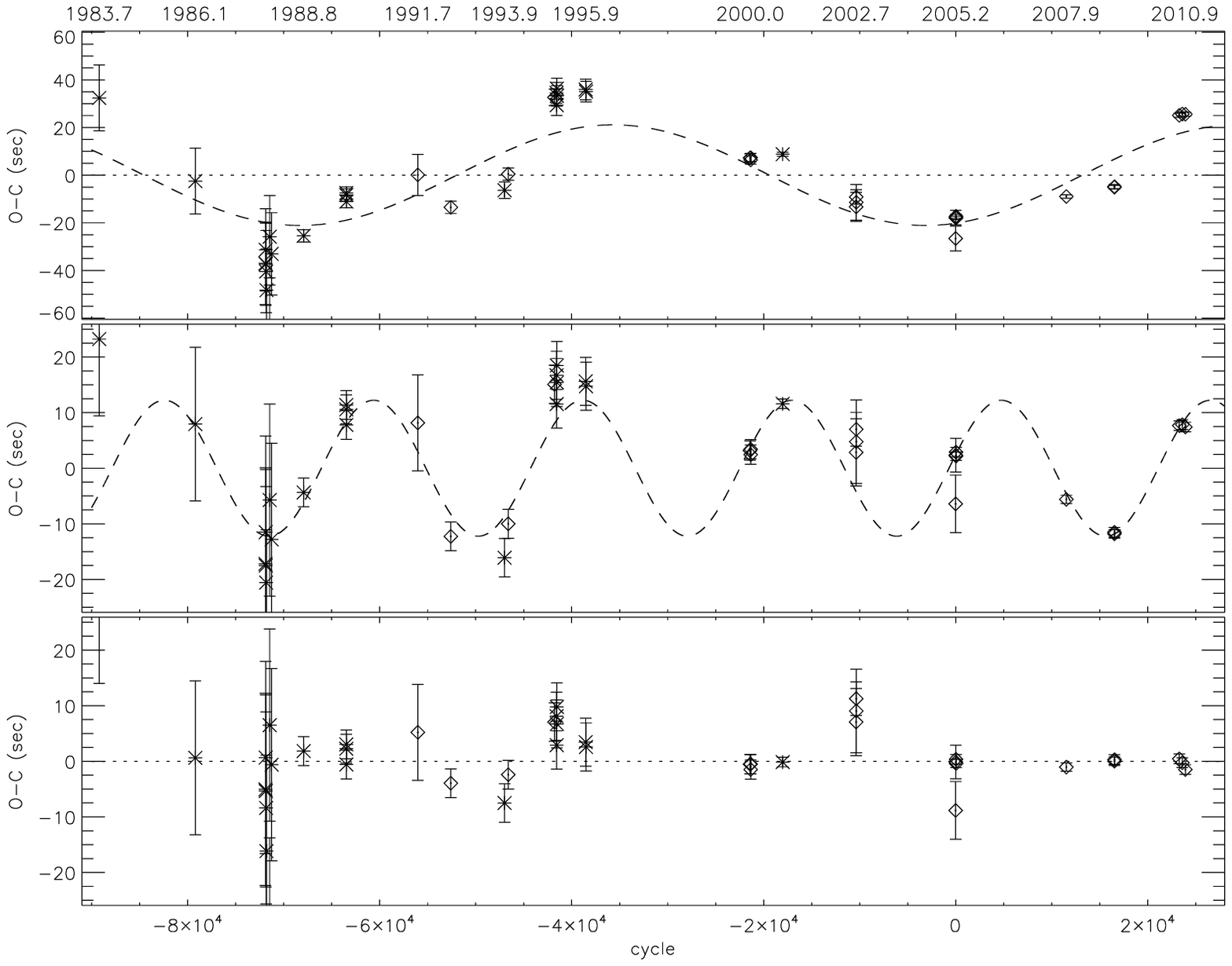} 
\caption{The O-C after successive subtraction of the three terms
  comprising our new eclipse ephemeris. Top: O-C after subtraction of
  the quadratic term with the first elliptical term overplotted
  (dashed curve). Middle: O-C after subtraction of the first
  elliptical term with the second elliptical term overplotted (dashed
  curve). Bottom: The final O-C residuals after subtraction of the
  second elliptical term. Diamonds are our new data or data
  that we have reduced from archives. Crosses are eclipse
  times from the literature and converted (by us) to BJD$_{TDB}$.}
\label{o-c}
\end{figure*}

\subsection{The eclipses}

A sample of our new and archival eclipse data is shown in
Fig~\ref{eclipses} phased on our new ephemeris (see below and table
\ref{tab:planets}). The eclipse profiles are of varying quality and at
multiple wavelengths. All of the eclipses can be understood in the
framework of the standard polar model and from the general literature
on UZ For (see section 1). UZ For undergoes periods of either one
(e.g.  Bailey \& Cropper 1991) or two-pole (e.g. Perryman et al. 2001)
accretion states, which has been most clearly captured by our SALTICAM
(2007) and BVIT (2009) observations respectively. During both
accretion states the rapid fall and rise in flux at phases
$\sim$-0.031 and $\sim$0.031 correspond to the ingress and egress of
the main accretion region. During the two-pole accretion state a
second accretion region is additionally present, seen as the fall and
rise in flux at phases $\sim$-0.027 and $\sim$0.027. The remaining
gradual fall and rise during phases $\sim -0.031$ to $-0.025$ and
$\sim 0.025$ to $0.031$ is attributed to the ingress and egress
respectively of the white dwarf photosphere and takes about 40s each.

\subsection{The O-C}

In table \ref{tab:observations} we list all of our new mid-eclipse
times as well as those we have measured from archival data or
extracted from the literature. The orbital period calculation of
Perryman et al. (2001) was used to calculate the cycle number for each
eclipse. The period is sufficiently accurate to unambiguously assign
cycle counts to the entire $\sim $ 27 years of eclipses. The eclipse
observed in our 2005 UCTPOL photometry was used to define the epoch
(cycle 0).  This period and epoch were next used as the starting point
to perform a least-square quadratic fit with appropriate weighting set
by the eclipse error measurements. The resulting fit gives a reduced
$\chi^{2}>95$ with peak-to-peak residuals of $\sim 60-80s$. Note that
we have not included the eclipses of Dai et al. (2010) in our analysis
as their O-Cs are over 300s compared to our quadratic ephemeris. We
believe either their measurement and/or time standard conversion to be
in error.

It is immediately apparent that there are significant departures from
the quadratic ephemeris with a trend that appears to be periodic (see top
plot of fig~\ref{o-c} for residuals, albeit from a different quadratic
fit: see below).

We next investigated the solutions resulting from models consisting of
a quadratic plus an elliptical fit to the eclipse times. A best
reduced $\chi^{2}$ of 6.2 was achieved. An F-test shows that
it is the better model (compared to the quadratic ephemeris) with a
confidence of $>99.999\%$ even though the elliptical term adds 5 more
parameters to the model.

However significant residuals still remain, ($\sim$10s) for some of
the eclipse times (not shown). We next attempted a simultaneous
quadratic plus two ellipticals fit to the eclipse times. The second
elliptical term adds a further five parameters to the model, giving 13
in total. Therefore, given the large number of parameters, a grid of
starting parameters for minimisation was required in order to explore
the resulting degeneracy in the solutions. Approximately $10^{7}$
minimisations were calculated. During minimisation all 13 parameters
were free to vary. Predictably, the results have better reduced
$\chi^{2}$ but with degeneracy in many of the parameters. Formally the
F-test confirms that adding a second elliptical term is the better
model with a high level of confidence ($>99.9999\%$) for the solutions
with reduced $\chi^{2}=1.0$. Other solutions with reduced
$\chi^{2}=4.0$ and $3.0$ are also significantly better with a 98\% and
99.9995\% confidence respectively.  We explore the degeneracy in the
multi-dimensional $\chi^{2}$ space in section 4.2.

\section{Discussion}

\subsection{The O-C}

% good grav rad and mag break discussion in Brinkworth

Our results suggest that the deviations in the eclipse O-C are best
described by the combination of a quadratic term plus two elliptical
terms. This is highly suggestive of both secular and cyclic period
variations.

Period changes in binary systems are generally understood to be due to
gravitational radiation, magnetic braking, solar-type magnetic cycles
in the secondary star (Applegate's mechanism) and/or the presence of a
third body in an orbit around the binary.

% Pdot = 2B/P = 2. * (-9.3122608e-14/0.0878654230) = -2.1196645e-12
%pdot err = 2. * sqrt((Berr/B)^2 + (Perr/P)^2) = 4.5e-14

%We do not detect any
%overall period decrease over the time-span of the observations which
%would be generally expected from magnetic braking and/or gravitational
%radiation (see e.g. Andronov et al. 2003 for an example of the
%application of these mechanisms to CVs). Instead, the period changes
%we observe in UZ For appear to be cyclic in nature.

%standard CV magnetic braking rate (Rappaport et al. 1983)

Applegate's mechanism and/or the presence of a third body would be
more consistent with cyclic variability. The latter would produce
strictly periodic cycles while non-strictly periodic cycles would be
expected from the former mechanism. Therefore, we next look at each of these mechanisms in turn.

%An increase in the confidence of the first sinusoid ($\sim $18.3
%years) would take a couple more cycles .... long (multiples of 18.3
%years) but a lot sooner to ``disprove''..... a longer basline is needed.

%Whatever the outcome of future observations, the fact remains that the
%observed O-C are real and intrinsic to the UZ For system.

\subsection{Tertiary and quaternary components}

\begin{figure}
\epsfxsize=8cm
\epsffile{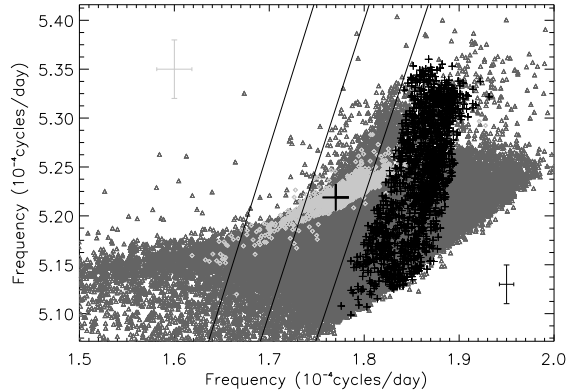} 
\caption{The reduced $\chi^{2}$ parameter space for the two elliptical
  periods. Black crosses, dark grey triangles and light grey diamonds
  are the solutions with reduced $\chi^{2}<1.0, 1.0<\chi^{2}<2.5$ and
  $1.0<\chi^{2}<2.5$ respectively. The diamonds have the additional
  constraint of both eccentricities $<0.1$. The large black cross
  represents the location of the solution shown in Fig ~\ref{o-c} with
  parameters listed in table~\ref{tab:planets}. From left to right,
  the diagonal lines represent contours of constant period ratios of
  3.1, 3.0 and 2.9. Typical one sigma errors are shown in the top left
  and bottom right corners.}
\label{per_per}
\end{figure}

%print,(1.D/(1.5e-4))/365.25D
%       18.252338
%IDL> print,(1.D/(2.0e-4))/365.25D
%       13.689254

%IDL> print,(1.D/(5.1e-4))/365.25D
%       5.3683351
%IDL> print,(1.D/(5.4e-4))/365.25D
%       5.0700942

We now explore the degeneracy in the multi-dimensional $\chi^{2}$
space of the model fits containing one quadratic plus two elliptical
terms in the context that the variations are due to the effect of
third and fourth bodies in the system. Then the changes in the O-C
arise because of the light-time effect caused by the gravitational
influence of the additional bodies.

As a first step, we plot the distribution of the period of the two
elliptical orbits for those solutions which had reduced $\chi^{2} <
2.5$ (Fig~\ref{per_per}). The starting grid, for minimisation, had
period values between 2 and 50 years and Fig~\ref{per_per} shows that
the minimised solutions have clustered in the period ranges $\sim
13-19$ and $\sim 5-5.5$ years.

All of the solutions with reduced $\chi^{2} \leq 1.0$ ($ \chi^{2} \leq
28$, shown as the black crosses in Fig~\ref{per_per}) show periods
centered on approximately 5.3 and 15 years, giving a period ratio of
$Rp\sim 2.8$ . The corresponding predicted eccentricities of these
solutions are $>>0.1$ for both ellipses. We performed N-body
simulations on a sample of these solutions. The Euler method was used
with sufficiently small time steps to ensure accurate calculations. We
tested the accuracy of our code by first applying it to single
elliptical orbits comparable to the innermost high eccentricity
orbit. Our code preserves the eccentricity and the semi-major axis to
better than 10\%, and the periastron angle to $<$ 0.1 radians over a
time period corresponding to $>10^{5}$ orbits of the outermost
body. We then defined orbital solutions that have essentially the same
two planet orbital elements to the starting conditions after $>10^{5}$
orbits of the longer period as stable. As expected, we found that they
all are very unstable orbits and are therefore unrealistic solutions.

We next looked at the solutions with reduced $\chi^{2}$ in the range
$1.0<\chi^{2}<2.5$, which are shown as the dark grey triangles in
Fig~\ref{per_per}. They occupy a larger area of the plot which
overlaps with the previous solutions. However most of these solutions
still require the longer period elliptical to have a large
eccentricity ($>0.1$ and typically 0.4). N-body simulations within
this parameter range also revealed unstable orbits.

%We therefore identified the solutions with eccentricities $<0.1$ for
%both planets, where our N-body simulations show stable orbits. These
%are represented as the light grey diamonds in Fig~\ref{per_per} which
%also overlap with the higher eccentric solutions presented above. The best
%of these more realistic solutions has a reduced $\chi^{2}=2.06$. The
%corresponding $\chi^{2}\sim 56$ (41-13-1 degrees of freedom) is below
%the conventionally accepted significance level of $\sim 0.05$.

We therefore identified those solutions which had orbital
eccentricities $<$ 0.1 for both planets and which had stable orbits
according to our N-body simulations. These are represented as the
light grey diamonds in Fig~\ref{per_per} which also overlap with the
higher eccentric solutions presented above. However, the best of these
solutions give ${\chi}^{2}$=58 (28 dof, giving
{${\chi}_{\nu}^{2}$}=2.06), which is a significantly poorer fit
compared to our best-fit highly eccentric orbital solutions and also
gives a formally poor fit to our data.

We note that our models assume the two planets to be co-planar. There
may yet be a set of realistic, more eccentric, solutions if the
planets have inclined orbits with respect to each other. We have not
investigated this additional parameter space as our data set is not of
sufficient quality nor quantity to warrant it.

Given these caveats, in order to calculate the implied two planet
parameters, we selected the best-fit low eccentric and stable
solutions. Additionally we applied period errors that encompass the
whole range of solutions in Fig~\ref{per_per}. We note that the two
planet parameters calculated below are not specific to this one
best-fit solution but are representative of all the solutions with
reduced ${\chi}^{2} < 2.5$ shown in Fig~\ref{per_per}. The
calculations are independent of the eccentricities. This particular
solution is marked as the large black cross in Fig~\ref{per_per}, the
two elliptical parameters are listed in table~\ref{tab:planets} and
over- plotted on the O-C in Fig~\ref{o-c}.

% $(0.7-1.1)\times 10^{-4}$ and $(2.75-3.0)\times 10^{-4}$.

The amplitudes of the oscillations can be used to calculate the
projected distances $a sin(i)$ from the center of mass of the binary
to the center of mass of each of the triple systems (0.042(1) and
0.025(1) AU for the long and short respectively). Setting the binary
mass to be $0.7 M_{\odot}+0.14 M_{\odot}$ gives the corresponding mass
functions ($f(m_{3,4}) = 2.9 \times 10^{-7} M_{\odot}, 5.3 \times
10^{-7} M_{\odot})$. With the binary inclination at $i=80^{\rm o}$ the
respective minimum masses for the third and fourth bodies (assuming
they are in the plane of the binary) are 0.006(1) $M_{\odot}$ and
0.007(1) $M_{\odot}$ and would therefore qualify as extrasolar giant
planets (6.3(1.5)$M_{Jup}$ and 7.7(1.2)$M_{Jup}$) for orbital
inclinations $i_{3} > 25^{\rm o}, i_{4} > 32^{\rm o} $ respectively.
The quoted errors include the range in periods shown in
Fig~\ref{per_per} and not the formal one sigma errors of one of the
solutions. In addition, the quoted errors include the propagated
uncertainties in the inclination and binary mass.  These parameters
are summarised in table \ref{tab:planets}.

%We note
%that the semi-major axis of even the shortest period object poses no
%problem for orbit stability (Holman \& Wiegert 1999).

The equation for the times ($T$) of mid-eclipse of the main accretion
spot are then given by:

\begin{eqnarray*}
T(BJD_{TDB}) &=& T_{0} + P_{bin}E  + AE^{2}\\
    &+& K_{bin,(3)} sin(\upsilon_{3}- \varpi_{3}) {(1 - e_{(3)}^{2}) \over (1 + e_{(3)}cos(\upsilon_{3})}  \\
    &+& K_{bin,(4)} sin(\upsilon_{4}- \varpi_{4}) {(1 - e_{(4)}^{2}) \over (1 + e_{(4)}cos(\upsilon_{4})}  \\
\end{eqnarray*}

$T_{0}, P_{bin}, A, E$ are the time of epoch, the binary orbital
period (days), the quadratic parameter (related to the rate of period
decrease by $\dot{P}_{bin} = 2A/P_{bin}$) and the binary cycle number
which comprise the quadratic term of the ephemeris. In the context
that the two elliptical terms are due to third and fourth bodies in
the system, then the parameters of the elliptical terms are:
$K_{bin,(3,4)}$ are the amplitudes of the eclipse time variations as a
result of the light-travel-time effect of the two bodies,
$\upsilon_{(3,4)}$ are the true anomalies of the two bodies, which
progresses through $2\pi$ over the orbital periods ($P_{(3,4)}$) and
are functions of $E$, the times of the periastron passages
($T_{(3,4)}$) and the orbital frequencies (in cycles per binary
orbital cycle) of the two bodies ($f_{3,4}$). $e_{(3,4)}$ are the
eccentricities and $\varpi_{(3,4)}$ are the longitudes of periastron
passage measured from the ascending node in the plane of the
sky. Similar elliptical variations have been seen in NN Ser and DP Leo
(Beuermann et al 2010a,b).

Fig~\ref{o-c} shows the fit and the O-C residuals after successive
subtraction of the three terms comprising our new eclipse ephemeris
with parameters listed in table \ref{tab:planets}. The top and middle
plots show that the two elliptical terms describe the time of eclipse
variations very well. The lower plot shows the final O-C residuals
after subtraction of the full ephemeris. Some residuals still remain
which could be reduced further if larger eccentricities were
permitted, particularly for the outer planet. Such orbital solutions
may exist, especially given the possible indication that the planets
could be locked in a 3:1 ratio. However better sampled observations
are required to further constrain the solutions.

%The residuals are clearly
%minimal with a final reduced $\chi^{2}=0.87$ for this particular
%solution.

%In fact, as can be seen from fig ...

%The observed mass function (Borkovits \& Hegedues 1996) is 1.1 × 10− 7
%Mimplying a minimum mass of the third body of 0.0047 M (5 Jupiter
%masses) for an assumed total mass of the system of 1.0 M ⊙ . On the
%other hand, a third body more massive than an object at the
%hydrogen-burning limit would require an inclination <3 and is
%therefore not very likely.

\subsection{The secular variability}

The secular variability amounts to a decrease in the orbital period of
$\dot{P}_{bin} = -1.56(5)\ 10^{-12} s\ s^{-1}$. Of the 1416 solutions
that comprise the `chosen' parameter space (light grey area in
Fig~\ref{per_per}) only one solution showed a $\dot{P}_{bin}$
consistent with 0. It has a reduced $\chi^{2}=2.6$. The rest predict a
minimum in the period decrease rate of $\dot{P}_{bin} = -1.0\ 10^{-12}
s\ s^{-1}$. Formally an F-test shows that adding the quadratic
parameter to the ephemeris is the better model with a 99.99\% level of
confidence ($\chi^{2}=109, 86$ for $30, 29$ degrees of freedom for the
two models respectively).

Similar levels of period decrease has been detected in other similar
short period binaries e.g. DP Leo (Schwope et al. 2002), NN Ser
(Brinkworth et al. 2006) and HU Aqr (Schwarz et al. 2009) for which
gravitational radiation and magnetic braking have been shown to be
either insufficient or problematic in the framework of the standard CV
evolutionary model. Future observations are needed to show if these
variations are indeed secular or periodic.

%The implied loss of angular momentum could well be explained by a
%combination of gravitational radiation and magnetic braking
%(Rappaport et al. 1983).

\subsection{Applegate mechanism}

\begin{figure}
\epsfxsize=8cm
\epsffile{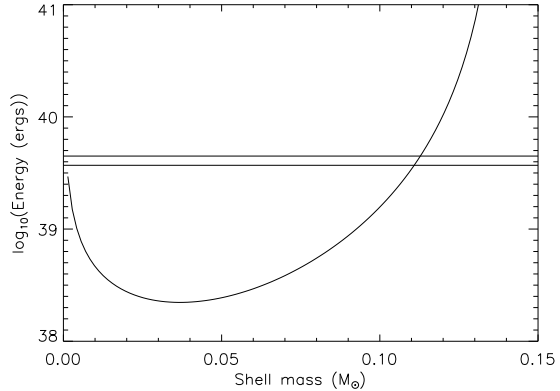} 
\caption{Solid curve shows the energy required to effect the period
  change observed in UZ For as a function of assumed shell mass, using
  Applegate's (1992) mechanism. The two horizontal lines represent the
  total radiant energy of the secondary (assuming $2880 < T_{eff} <
  3020$K) and hence the amount of energy available. }
\label{applegate}
\end{figure}

Applegate (1992) proposed that solar-like magnetic cycles would drive
shape changes in the secondary, thus redistributing the angular
momentum within the star, changing its oblateness. This then leads to
a change in its quadrupole moment and consequently a change in orbital
period at the expense of some energy.  This has been the preferred
mechanism to explain such cyclic variations in CVs and other long
period close binaries (Algol, RS CVn and W UMa stars).

Following the prescription of Applegate (1992), the energy required to
generate a period change is:
$$ \Delta E = \Omega_{dr}\Delta J + {\Delta J^{2} \over 2I_{eff}}$$
The initial differential rotation $\Omega_{dr}\Delta J$ is set to zero
in order to calculate the minimum energy required. The effective
moment of inertia $I_{eff} = I_{S}I_{*}/(I_{S}+I_{*})$ is calculated
by assuming the star is separated into a shell $I_{S}$ and a core
$I_{*}$. We experimented with a range of shell masses. $\Delta J$ is
the change in angular momentum and is given by
$$\Delta J = {-GM^{2} \over R} {\left( a \over R \right) }^{2} {\Delta
  P \over 6 \pi}$$ We used the secondary star mass from Bailey \&
Cropper (1991) namely $M = 0.14M_{\odot}$ and the corresponding radius
$R = 0.177 R_{\odot}$ following Patterson (1984). $a = 5.5\times10^8 m$ is the
binary separation using $a = 3.53 \times 10^{10} (M_{1}/M_{\odot})^{1/3} (1 + q
)^{1/3} P^{2/3}_{orb} (h) $ (Warner 1995, equation 2.1b). $\Delta P$
can be obtained from equation (38) of Applegate (1992) relating the
amplitude of orbital period modulation and the amplitude of the O-C
oscillation:
$${\Delta P \over P} = 2 \pi {O-C \over P_{mod}} $$ where $P$ and
$P_{mod}$ are the orbital and modulation period respectively (using
$P_{3}$ listed in table \ref{tab:planets}). The solid curve in
Fig.~\ref{applegate} shows the minimum energy required to drive the
maximum observed period change in UZ For, as a function of assumed
secondary shell mass. The two horizontal lines represent the total
radiant energy of the secondary $L = 4 \pi R^{2} \sigma T^{4}$ over
the modulation period, assuming $2880 < T_{eff} < 3020$K, which
appears to be more than sufficient to drive the Applegate
mechanism. The situation is not so clear cut if one instead integrates
over shells and allows for the quadrupole moment of the core (using
the calculations of Brinkworth et al. 2006). This raises the minimum
energy by about an order of magnitude, which makes it comparable, at
minimum, to the energy of the star.

We should add that Lanza et al. (1998) propose a prescription that is
more energy efficient than the Applegate mechanism, perhaps by a
factor of two. Therefore, with further refinements, magnetic fields
may yet be shown to able to drive the period changes seen here.

%\subsection{Asynchronous White Dwarf}

%Some polars are thought to contain a asynchronous white dwarf
%(e.g. XXXX).

\subsection{Spot motion}

%It is clear from the top panel of Fig.~\ref{o-c} that there are large
%departures in the o-c values ($\sim 40-60s$) from a simple linear
%ephemeris for the orbital period. 

The eclipse times are derived from the observed ingress and egress
times of the accretion spot and not the center of the white dwarf
itself. Therefore the observed O-Cs could be as a result of motion of
the spot on the white dwarf.  In addition, the egress and ingress of
the white dwarf photosphere has been observed to take about 40 seconds
(seen unambiguously in the low state observations of Bailey \& Cropper
(1991) and confirmed in our low state {\it SALTICAM 2007}
observations) and therefore could accommodate a $\sim 40-60s$ of
spot motion.

To assess this possibility we investigated the actual morphologies of
the eclipse profiles in order to measure the relative phases of the
white dwarf photosphere ingresses and egresses to that of the main
accretion spot. Accordingly, we display a sample of our eclipse
observations in Fig.~\ref{eclipses} phase folded and binned on our new
ephemeris. The upper two eclipse profiles are the {\it HIPPO 2010} and
the {\it BVIT 2009} observations respectively. These correspond to the
two newest data points in Fig.~\ref{o-c} which show a O-C shift of
$\sim$ 30s with respect to each other on the quadratic subtracted O-C
plot (upper plot Fig.~\ref{o-c}). Therefore if the spot did indeed
change locations, on the surface of the white dwarf, between these two
observations then we should observe a corresponding relative phase
shift between spot ingress/egress to that of the white dwarf
photosphere ingress/egress. However one can see that the beginning of
the white dwarf photosphere egress (solid vertical line at phase $\sim
0.026$) is consistently $\sim 40s $ (0.005 phase) ahead of the spot
egress (vertical grey bars) between these two observations. Thus, one
would have expected the relative time difference between the white
dwarf photosphere and spot to be shorter by $\sim 30s$ between the two
observations and not approximately equal as observed. Therefore we
conclude that the spot has not moved on the surface of the white dwarf
during these observations, at least within our measurable errors of
about 1-5s. The same relative phase difference (spot, white dwarf
photosphere) is also apparent in the other eclipse profiles in which
the spot and white dwarf photosphere are resolved (see the next five
plots in Fig.~\ref{eclipses} corresponding to the {\it SALTICAM 2007,
  UCTPOL 2005, SWIFT 2002, UCTPOL 2002} and {\it SCAM 1999}
observations). Furthermore, the same unchanging relative time
differences are seen in the ingresses, although the white dwarf
photosphere ingress emission may be complicated by an additional
contribution from the accretion stream. The stream is not visible
during the white dwarf photosphere egress, which can be understood
from simple eclipse geometrical arguments, and confirmed through {\it
  HST} UV spectroscopy (Stockman \& Schmidt 1996).

We note, however, that the longitude of the accretion spot should be
expected to change during different accretion states. For example,
Schwope et al. (2001) calculated a change in spot longitude of $\sim
10^{\rm o}$ between high and intermediate accretion states in the
eclipsing polar HU Aqr. This would translate to about a shift of $\sim
2-3s$ in the O-C values (Schwarz et al. 2009) which therefore cannot
account for the observed O-C values. This implies that if there was a
similar spot motion in UZ For during different accretion states, it
cannot account for the large shift seen in the measured O-C values.

Additionally there was not any measurable movement of the spot in
latitude during our observations. This is apparent from
table~\ref{tab:observations} where the eclipse width measurements
agree within errors: a change in spot latitude would have resulted in
a corresponding change in eclipse length.

\section{Summary and Conclusion}

We have detected departures in the eclipse times of UZ For from a
simple quadratic ephemeris of up to $\sim$60s. The departures are
suggestive of two periodicities of $\sim$16 and $\sim$5.25 years. The
two favoured mechanisms to drive the periodicities are either two
giant extrasolar planets as companions to the binary or a magnetic
cycle mechanism (e.g. Applegate's mechanism) of the secondary
star. However, Applegate's mechanism would require the entire radiant
energy output of the secondary and therefore would seem to be the
least likely of the two, barring any further refinements in the effect
of magnetic fields (e.g. Lanza 1998). A two planet model is also
problematic given the quality of the data in that a high eccentric
orbit, for at least one of the planets, seems to be required to fully
capture all of the eclipse times.

%The strong indication of a 3:1 period ratio
%would seem to add more weight to the two planet model.

%As yet there is insufficient data to favour between the two mechanisms
%and indeed both could be present, although presumably the Applegate
%mechanism would drive one of the periods only.  Future eclipse
%measurements on a relatively short timescale (a few years) would
%probably suffice to secure the $\sim$ 3 year periodicity.

If it can be confirmed that the residuals are due to a third and
a fourth body, then the planets either formed in a pre-common
envelope circumbinary protoplanetary disc (first generation) or in a
disc that resulted from the common envelope (CE) phase (second generation:
Perets 2010). The separation of the progenitor binary is of the order
of a few AU, comparable to that of the planets, which implies that only second generation planets could
have formed at the orbits suggested here. However, Beuermann et al (2010a)
suggested for the planets around NN Ser, a slowly expanding CE could
provide the dynamical force to drag inwards planets formed further out, which
would have otherwise been lost to the system due to the decrease in
mass of the central binary (Alexander et al. 1976). In either case, we
note that the semi-major axis of even the shortest period object poses
no problem for orbit stability (Holman \& Wiegert 1999).

%these results would provide further evidence that planets (and brown
%dwarfs - cf Maxted et al 2006) can survive a common envelope phase. Or
%perhaps suggest a post-CE origin for second generation planet
%formation. In the case of UZ For, our analysis suggests that the inner
%planet has a semi-major axis of only 2.8 AU. This suggests that a
%pre-common envelope origin to be very unlikely given the separation of
%the progenitor binary.

It is intriguing that Qian et al (2011) propose a very similar two
elliptical model fit for the polar HU Aqr, also using eclipse timing
results. In particular they also find that the larger ellipse requires
a high eccentricity (0.51) to correctly capture all of the data. Therefore, their two planet model for HU Aqr seems to have a similar problem with orbit instabilities that we have found for UZ For.

As yet there is insufficient data on UZ For to identify conclusively the mechanism responsible for the periodic changes in its eclipse times, and indeed more than one mechanism could be present. Further good
signal/noise, high time resolved observations of UZ For and other
similarly eclipsing systems are encouraged.

%The HU Aqr
%observations presented in Qian et al (2011) are of better quality and
%the high eccentricity for one of the ellipticals is convincingly
%required to give a good fit.

%Say something about bad data points.

%List of eclipses where there is good S/N to see photosphere of WD.
%HIPPO oct 2010 one pole
%HIPPO sept 2010 one pole  $11_12_both_chns.phase$
%HIPPO Jan 2009  and 2nd pole
%BVIT Jan 2009   and 2nd pole
%SALTICAM 2007 def only 1 pole
%UCTPOL 2005 3 eclipses .. best to use $plot_20050205_BJD$  2 poles
%UCTPOL 2002 20020807 see one pole  ... maybe 2nd
%SWIFT 2002 one pole
%XMM ON .... not sufficiant S/N
%SCAM 1999 2 poles
%EUVE 1993 and 1995  cant see photosphere of WD .... but def worth while including in plot to compare
%ROSAT cannot see photosphere ....as expected

\section{Acknowledgments}

We thank for referee for an insightful report that has significantly
improved the paper.

This material is based upon work supported financially by the National
Research Foundation.  Any opinions, findings and conclusions or
recommendations expressed in this material are those of the author(s)
and therefore the NRF does not accept any liability in regard thereto.

Some of the observations reported in this paper were obtained with the
Southern African Large Telescope (SALT).

Some of the data presented in this paper were obtained from the
Multi-mission Archive at the Space Telescope Science Institute
(MAST). STScI is operated by the Association of Universities for
Research in Astronomy, Inc., under NASA contract NAS5-26555. Support
for MAST for non-HST data is provided by the NASA Office of Space
Science via grant NAG5-7584 and by other grants and contracts.

We would also like to thank Drs. Jean Dupuis, Phil Hodge and Damian
J. Christian for their invaluable help when reducing EUVE and HST
archival data.

%Some of the observations reported in this paper were obtained with the
%Southern African Large Telescope (SALT), a consortium consisting of
%the National Research Foundation of South Africa, Nicholas Copernicus
%Astronomical Center of the Polish Academy of Sciences, Hobby Eberly
%Telescope Founding Institutions, Rutgers University,
%Georg-August-Universität Göttingen, University of Wisconsin - Madison,
%Carnegie Mellon University, University of Canterbury, United Kingdom
%SALT Consortium, University of North Carolina - Chapel Hill, Dartmouth
%College, American Museum of Natural History and the Inter-University
%Centre for Astronomy and Astrophysics, India.

\end{document}